\documentclass[12pt]{article}

\usepackage{graphicx}
\usepackage{axodraw}

\def\beq   {\begin{equation}}
\def\eeq   {\end{equation}}
\def\bea {\begin{eqnarray}}
\def\eea {\end{eqnarray}}

\def\dr{\ifmmode{\overline{\rm DR}} \else{$\overline{\rm DR}$} \fi}
\def\ms{\ifmmode{\overline{\rm MS}} \else{$\overline{\rm MS}$} \fi}

\begin{document}

\pagestyle{plain}

\vspace*{-1cm} 
\begin{flushright}
  TU-672 \\
\end{flushright}

\vspace*{0.7cm}

\begin{center}

{\Large \bf
Two-loop renormalization of $\tan\beta$ \\and its gauge dependence}
\footnote{Talk based on 
Ref.~\cite{mypaper} 
at SUSY02: the 10th International Conference on Supersymmetry 
and Unification of Fundamental Interactions, DESY, Hamburg, Germany, 
June 17--23, 2002.
}
\\

\vspace{10mm}

{\large Youichi Yamada}

\vspace{6mm}
\begin{tabular}{l}
{\it Department of Physics, Tohoku University, Sendai 980-8578, Japan}
\end{tabular}

\end{center}

\vspace{1cm}

\begin{abstract}
\baselineskip=15pt
Renormalization of two-loop divergent corrections to the vacuum 
expectation values ($v_1$, $v_2$) of the two Higgs doublets in the 
minimal supersymmetric standard model, 
and their ratio $\tan\beta=v_2/v_1$, is discussed 
for general $R_{\xi}$ gauge fixings. 
When the renormalized ($v_1$, $v_2$) are defined to give 
the minimum of the loop-corrected effective potential, it is shown that, 
beyond the one-loop level, the dimensionful parameters in the $R_{\xi}$ 
gauge fixing term generate gauge dependence of the renormalized $\tan\beta$. 
Additional shifts of the Higgs fields are necessary to realize 
the gauge-independent renormalization of $\tan\beta$. 
\end{abstract}

\vspace{15mm}

\baselineskip=15pt

The minimal supersymmetric (SUSY) standard model 
(MSSM) \cite{mssm,gh} has two Higgs boson doublets
\beq
H_1=(H_1^0, H_1^-),\;\;\; H_2=(H_2^+, H_2^0).  \label{eq1} 
\eeq
Both $H_1^0$ and $H_2^0$ acquire the vacuum expectation 
values (VEVs) $v_i$ ($i=1,2$) which spontaneously break 
the SU(2) $\times$ U(1) gauge symmetry. 
$H_i^0$ are then expanded about the minimum of the Higgs potential as 
\beq
H_i^0 = v_i/\sqrt{2}+\phi_i^0, \;\;\;
\langle \phi_i^0\rangle = 0.  \label{eq2}
\eeq
Here I ignore the CP violation in the Higgs sector 
and take $v_i$ as real and positive. 

A lot of physical quantities of the theory depend on the Higgs VEVs. 
In calculating radiative corrections to these quantities, $v_i$ 
have to be renormalized. In the MSSM, the renormalization is 
usually performed \cite{tanbrun,cpr} by specifying 
the weak boson masses, which are proportional to $v_1^2+v_2^2$, 
and the ratio $\tan\beta\equiv v_2/v_1$. However, since $\tan\beta$ itself 
is not a physical observable, a lot of renormalization 
schemes for $\tan\beta$ have been proposed \cite{dpf,freitas} 
in the studies of the radiative corrections in the MSSM. 
One method is to define the renormalized $\tan\beta$ as 
a tree-level function of the physical observables. This method is 
manifestly independent of the gauge fixing and renormalization scale. 
However, the form of the counterterm strongly depends on the chosen 
observables and is often very complicated. 
Here I concentrate on another method, the process-independent 
definition of $\tan\beta$, which is given by the ratio of the 
renormalized VEVs $v_i$. I discuss the renormalization of the 
ultraviolet (UV) divergent corrections to $v_i$ and $\tan\beta$, 
working in the \dr scheme \cite{DR}. The results are presented as 
the renormalization group equations (RGEs) for $v_i$ and $\tan\beta$. 
Since they are not physical observables, they may 
depend on the gauge fixing in general. I therefore 
investigate \cite{mypaper} their gauge dependence 
in the general $R_{\xi}$ gauge fixing \cite{rxi}. 
The results for the gauge dependence can be generalized 
for other models with two or more Higgs doublets. 

In the \dr scheme \cite{RGEvsusy,tanbrun,cpr}, we 
absorb $\Delta v_i$ by the shift of quadratic terms in the Higgs 
potential, as 
\beq
(m_i^2+\delta m_i^2)|\phi_i^0|^2 \;\Rightarrow \;
\delta L \ni -\sqrt{2}(v_i\delta m_i^2){\rm Re}\phi_i^0 .
\eeq
The renormalized $v_i$ then give the minimum of 
the loop-corrected effective potential $V_{\rm eff}(H_1, H_2)$. 
This scheme is very convenient in practical calculation, 
because of very simple counterterm for $\tan\beta$, and that 
the explicit forms of the tadpole diagrams are necessary only 
for two-point functions of the Higgs bosons. 
However, the effective potential is generally dependent on the 
gauge fixing \cite{veff}. 
The gauge dependence of the renormalized $v_i$ and 
their ratio $\tan\beta$ then might be a serious problem in 
calculating radiative corrections. 
I will therefore discuss the gauge dependence of the 
running $\tan\beta$ in this definition, in general $R_{\xi}$ gauges and 
to the two-loop order. 

The RGE for $v_i$ can be obtained from the UV divergent corrections to 
the two quark masses $m_b$ and $m_t$, ignoring the masses of all 
other quarks and leptons. 
These mass terms are generated from the $b\bar{b}H_1$ and $t\bar{t}H_2$ 
Yukawa couplings, respectively, as 
\beq
L_{\rm int} = -h_b \bar{b}_R b_L (v_1/\sqrt{2}+\phi_1^0)
-h_t \bar{t}_R t_L (v_2/\sqrt{2}+\phi_2^0) 
+ {\rm h.c.} \label{eq3}
\eeq
One then obtain 
\bea
\frac{dv_i}{dt} &=& 
\frac{1}{h_q}\left[ \sqrt{2}\frac{d}{dt}(m_q) -\frac{dh_q}{dt}v_i \right] , 
\label{eq8}
\eea
where $q=(b, t)$ for $i=(1,2)$, respectively. 
$t\equiv \ln Q_{\dr}$ is the \dr renormalization scale. 

The $R_{\xi}$ gauge fixing term takes the form 
\bea
L_{GF} &=& 
-\frac{1}{2\xi_Z} (\partial^{\mu}Z_{\mu} - \rho_Z G_Z)^2 
-\frac{1}{\xi_W} |\partial^{\mu}W^+_{\mu} -i\rho_W G_W^+|^2 \nonumber \\
&& -\frac{1}{2\xi_{\gamma}} (\partial^{\mu}\gamma_{\mu})^2
-\frac{1}{2\xi_g}\sum_{a=1}^8 (\partial^{\mu}g^a_{\mu})^2. \label{eq4}
\eea
The would-be Nambu-Goldstone bosons $G_V$ for $V=(Z,W)$ appear in 
Eq.~(\ref{eq4}). 
The parameters $\rho_V\equiv \xi_Vm_V$, where $m_V^2=g_V^2(v_1^2+v_2^2)/4$ 
($g_W^2=g_2^2$, $g_Z^2=g_2^2+g_Y^2$) are masses of $Z$ and $W^{\pm}$, 
are introduced in Eq.~(\ref{eq4}). 
This is to emphasize that the gauge symmetry 
breaking terms $\xi_Vm_V$ in $L_{GF}$, and also 
in the accompanied Fadeev-Popov ghost term, 
has very different nature from $v_i$ generated by the shifts (\ref{eq2}), 
as shown later. 
The terms $\rho_VG_V$ in Eq.~(\ref{eq4}) are expressed 
in the gauge basis (\ref{eq1}) of the Higgs bosons as 
\beq
\rho_Z G_Z = \xi_Z m_Z G_Z \equiv -{\sqrt{2}}{\rm Im} 
(\rho_{1Z} \phi_1^0 - \rho_{2Z} \phi_2^0), \label{eq5}
\eeq 
\beq
\rho_W G_W^{\pm} = \xi_W m_W G_W^{\pm} \equiv 
-(\rho_{1W} H_1^{\pm} - \rho_{2W} H_2^{\pm}), \label{eq6}
\eeq
with parameters $\rho_{iV}$. 
The usual form of the $R_{\xi}$ gauge fixing in the MSSM is 
recovered by the substitution \cite{cpr,gh} 
\beq
(\rho_{1V}, \rho_{2V})=\xi_Vg_V(v_1, v_2)/2=
\xi_Vm_V(\cos\beta, \sin\beta). \label{eq7}
\eeq

The UV divergent corrections to $m_b$ contain one source for  
the SU(2)$\times$U(1) gauge symmetry breaking. It is either 
$v_1$ originated from the shift (\ref{eq2}) of $H_1^0$, 
or $\rho_{1V}$ in the $R_{\xi}$ 
gauge fixing term (\ref{eq4}) and the Fadeev-Popov ghost term. 
The former contribution is obtained from that to the 
$\bar{b}_Rb_L\phi_1^0$ Yukawa coupling $h_b$ by replacing 
external $\phi_1^0$ by $v_1/\sqrt{2}$, 
except for the wave function correction of $H_1^0$ to $h_b$. 
Similar argument holds for the UV divergent corrections to $m_t$ and 
to the $\bar{t}_Rt_L\phi_2^0$ Yukawa coupling $h_t$. As a result, 
if the $\rho_{iV}$ contributions are absent, 
the runnings of $v_i$ are the same as those of the wave functions 
of $H_i^0$, namely 
\beq
\frac{dv_i}{dt}= -\gamma_i v_i . \label{naive}
\eeq
The anomalous dimensions $\gamma_i$ of $H_i^0$ generally depend on 
the gauge fixing parameters $\xi$. Their explicit forms are 
\beq
(4\pi)^2\gamma_i^{(1)} = N_c h_q^2 
-\frac{3}{4}g_2^2\left( 1-\frac{2}{3}\xi_W-\frac{1}{3}\xi_Z \right) 
-\frac{1}{4}g_Y^2(1-\xi_Z), \label{eq12}
\eeq
at the one-loop, and 
\bea
(4\pi)^4\gamma_1^{(2)} &=& -N_c(3h_b^4 + h_b^2h_t^2) 
 +2N_c h_b^2\left( \frac{8}{3}g_3^2 - \frac{1}{9} g_Y^2 \right) +L(g), 
\nonumber \\
(4\pi)^4\gamma_2^{(2)} &=& -N_c(3h_t^4 + h_b^2h_t^2) 
 +2N_c h_t^2\left( \frac{8}{3}g_3^2 + \frac{2}{9} g_Y^2 \right) +L(g). 
\label{eq13}
\eea
at the two-loop. Here $h_q^2=(h_b^2,h_t^2)$ for $i=(1,2)$, respectively, 
and $N_c=3$. The results in Eq.~(\ref{eq13}) are obtained from the general 
formula \cite{2loopNS} in the \ms scheme, after conversion 
into the \dr scheme \cite{mstodr}. Their last term $L(g)$ is a 
gauge-dependent ${\cal O}(g^4)$ polynomial 
and is common both for $\gamma_1^{(2)}$ and $\gamma_2^{(2)}$, while 
the ${\cal O}(h_q^2g^2)$ terms are $\xi$ independent \cite{2loopNS}. 
It is therefore seen that the gauge dependence of $\gamma_i$ cancels 
in the RGE of the ratio $\tan\beta$ to the two-loop order. 

However, in general $R_{\xi}$ gauges, $\rho_{iV}$ in the gauge fixing 
terms (\ref{eq4}) may give additional contributions to 
the quark mass running, as 
$\bar{b}b\rho_{1V}$ and $\bar{t}t\rho_{2V}$. Since they have 
no corresponding contributions to the 
$\bar{b}b\phi_1$ and $\bar{t}t\phi_2$ 
couplings, the RGEs for $v_i$ deviate \cite{schilling,cpr} from 
Eq.~(\ref{eq8}), as shown in Fig.~1 at the one-loop level. 
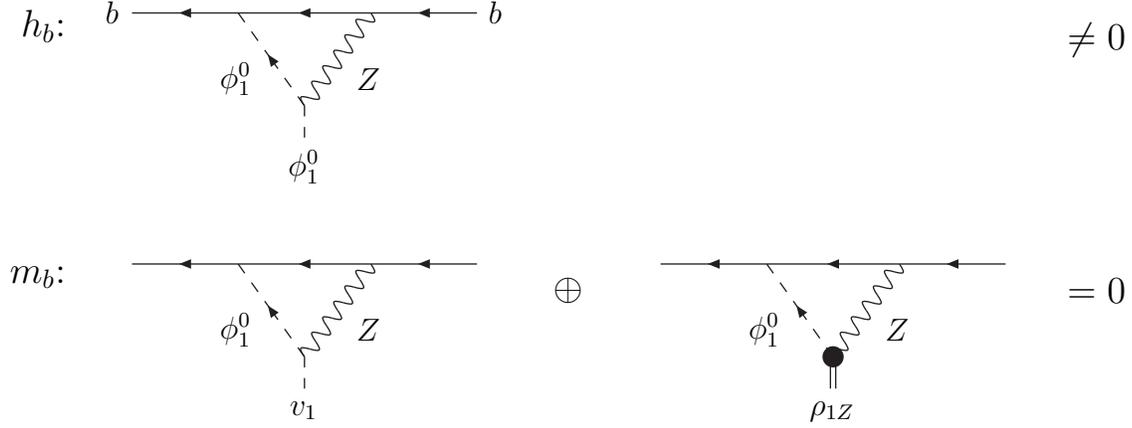
\begin{figure}[htbp]
{
\begin{center}
\begin{picture}(600,170)(0,0)

\SetOffset(40,165)

\Text(-10,-5)[r]{\large $h_b$:}
\Text(10,0)[r]{$b$}
\ArrowLine(55,0)(15,0)
\ArrowLine(105,0)(55,0)
\ArrowLine(145,0)(105,0)
\Text(150,0)[l]{$b$}
\DashArrowLine(80,-35)(55,0){5}
\Text(60,-25)[r]{$\phi_1^0$}
\Photon(80,-35)(105,0){3}{6}
\Text(100,-25)[l]{$Z$}
\DashLine(80,-47)(80,-35){5}
\Text(80,-52)[t]{$\phi_1^0$}

\SetOffset(240,165)


\Text(180,-10)[]{\large $\neq0$}

\SetOffset(40,70)

\Text(-10,-5)[r]{\large $m_b$:}
\ArrowLine(55,0)(15,0)
\ArrowLine(105,0)(55,0)
\ArrowLine(145,0)(105,0)
\DashArrowLine(80,-35)(55,0){5}
\Text(60,-25)[r]{$\phi_1^0$}
\Photon(80,-35)(105,0){3}{6}
\Text(100,-25)[l]{$Z$}
\DashLine(80,-47)(80,-35){5}
\Text(80,-52)[t]{$v_1$}

\Text(180,-10)[]{\large $\oplus$}

\SetOffset(240,70)

\ArrowLine(55,0)(15,0)
\ArrowLine(105,0)(55,0)
\ArrowLine(145,0)(105,0)
\DashArrowLine(80,-35)(55,0){5}
\Text(60,-25)[r]{$\phi_1^0$}
\Photon(80,-35)(105,0){3}{6}
\Text(100,-25)[l]{$Z$}
\Line(79,-35)(79,-47)
\Line(81,-35)(81,-47)
\Vertex(80,-35){4}
\Text(80,-52)[t]{$\rho_{1Z}$}

\Text(180,-10)[]{\large $=0$}

\end{picture}
\end{center}
}
\label{fig1}
\caption{ 
One-loop difference between the runnings of $h_b$ and $m_b$ by 
the $\rho_{1Z}$ contribution.}
\end{figure}
The general forms of the RGEs are then 
\beq
\frac{dv_i}{dt}= -\gamma_i v_i + Y_{iV} \rho_{iV}, \label{eq9}
\eeq
where $Y_{iV}$ are polynomials of dimensionless couplings. 
Therefore, the RGE for $\tan\beta$ becomes, substituting Eq.~(\ref{eq7}), 
\beq
\frac{d}{dt}\tan\beta =\tan\beta \left( -\gamma_2 +\gamma_1 
+ \frac{\xi_V g_V}{2}(Y_{2V} - Y_{1V})
\right) . \label{eq10}
\eeq 

I then give explicit form of the RGE for $\tan\beta$ to the 
two-loop order. First, one-loop RGEs for $v_i$ ($i=1,2$) are 
\bea
\left. 
\frac{dv_i}{dt}\right|_{\rm 1loop} &=& 
-\gamma_i^{(1)}v_i +\frac{1}{(4\pi)^2}(g_Z \rho_{iZ} +2g_2 \rho_{iW}) 
\nonumber \\
&=& v_i \left[ -\gamma_i^{(1)} +\frac{1}{(4\pi)^2}\left(
\frac{\xi_Z g_Z^2}{2} + \xi_W g_2^2 \right) \right]\, , \label{eq11}
\eea
The $\rho_{iV}$ contributions to $m_q$ are obtained from 
the diagrams similar to that in Fig.~1. Eq.~(\ref{eq11}) is consistent 
with the result in Refs.~\cite{cpr} for $\xi=1$. 
Since the gauge dependence of $\gamma_i$, as well as the 
contribution from ($\rho_{iZ}$, $\rho_{iW}$) satisfying Eq.~(\ref{eq7}), 
cancels in the ratio (\ref{eq10}), the one-loop running $\tan\beta$ is 
gauge parameter independent in the $R_{\xi}$ gauge. 
Note that this one-loop gauge independence of the running $\tan\beta$ 
does not hold in more general gauge fixings where Eq.~(\ref{eq7}) is 
not satisfied \cite{freitas}. 

The two-loop $\rho_{iV}$ contributions to $dv_i/dt$ have 
${\cal O}(h_q^2g\rho_{iV})$ and ${\cal O}(g^3\rho_{iV})$ terms. 
The latter is common for 
both $i=1$ and 2, and cancels out in the ratio $\tan\beta$ if 
Eq.~(\ref{eq7}) is satisfied. Therefore, only the former 
${\cal O}(h_q^2g\rho_{iV})$ contributions are explicitly calculated. 
For example, the ${\cal O}(h_b^2g_Z\rho_{1Z})$ contribution to $v_1$ 
comes from the diagram in Fig.~2. 
\begin{figure}[htbp]
\begin{center}
\begin{picture}(450,160)(0,0)


\SetOffset(110,120)

\Text(0,0)[r]{$b$}
\ArrowLine(45,0)(5,0)
\ArrowLine(135,0)(45,0)
\Text(90,10)[b]{$b$}
\ArrowLine(175,0)(135,0)
\Text(180,0)[l]{$b$}
\DashArrowLine(90,-90)(70,-51){5}
\DashArrowLine(60,-29)(45,0){5}
\Text(39,-25)[r]{$\phi_1^0$}
\ArrowArc(65,-40)(12,117,297)
\ArrowArcn(65,-40)(12,117,-63)
\Text(52,-53)[r]{$b$}
\Photon(90,-90)(135,0){3}{9}
\Text(130,-40)[l]{$Z_L$}
\Line(89,-90)(89,-102)
\Line(91,-90)(91,-102)
\Vertex(90,-90){5}
\Text(90,-107)[t]{$\rho_{1Z}$}

\end{picture}
\end{center}
\caption{ 
Two-loop divergent ${\cal O}(h_b^2g\rho_{1Z})$ contribution to $m_b$. 
}
\label{fig2}
\end{figure}
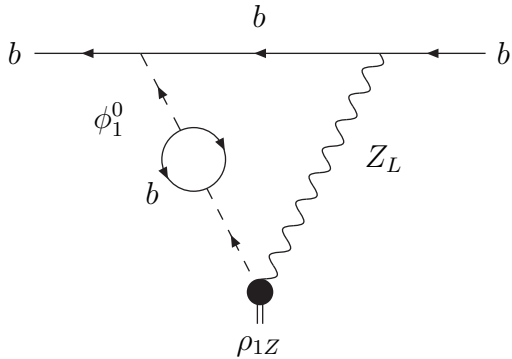
The two-loop RGEs for $v_i$ are finally 
\beq
\left. 
\frac{dv_i}{dt}\right|_{\rm 2loop} = 
-\gamma_i^{(2)} v_i 
- \frac{N_ch_q^2}{(4\pi)^4} (g_Z\rho_{iZ} + 2 g_W \rho_{iW})+P_V(g)\rho_{iV}, 
\label{eq14}
\eeq
where again $h_q^2=(h_b^2,h_t^2)$ for $i=(1,2)$, respectively. 
$P_V(g)$ are possibly gauge-dependent ${\cal O}(g^3)$ functions 
which are common for both $\rho_{1V}$ and $\rho_{2V}$. 
It is therefore seen that, due to the $\rho_{iV}$ contributions 
in Eq.~(\ref{eq14}), the running $\tan\beta$ has 
the ${\cal O}(h_q^2g_2^2,h_q^2g_Y^2)$ gauge parameter dependence. 
Although existing higher-order calculations of the corrections 
to the MSSM Higgs sector \cite{higgs2lEP,higgs2lRG,higgs2lFD,higgs2lCO} 
have not included the contributions of these orders yet, the gauge 
dependence of $\tan\beta$ may cause theoretical problem in future studies 
of the higher-order corrections in the MSSM. 

A possible way to restore the gauge independence of 
renormalized running $\tan\beta$ is to introduce gauge-dependent 
shifts of $\phi_i^0$ such as to cancel the $\rho_{iV}$ contributions 
to the effective action. 
This modification corresponds to the addition of extra shifts of $v_i$ 
to all diagrams. 
The running $v_i$ in this new definition then obey the same RGEs as 
those for $H_i$, namely Eq.~(\ref{naive}). 
The modified renormalized 
$\tan\beta$ becomes gauge independent to the two-loop order. 
However, an extra two-loop shift $\delta(v_2/v_1)$ has to be added to 
any quantities which depend on $\tan\beta$. The concrete procedure 
for this modification is now under investigation.

In conclusion, I discussed the two-loop UV renormalization 
of the ratio $\tan\beta=v_2/v_1$ of the Higgs VEVs in the MSSM 
in general $R_{\xi}$ gauges. 
When renormalized $v_i$ are given by the minimum of the 
loop-corrected effective potential, the contributions of $\rho_{iV}$ in 
the $R_{\xi}$ gauge fixing term cause two-loop gauge dependence 
of the RGE for $\tan\beta$. To avoid this gauge dependence, 
the contributions of $\rho_{iV}$ have to be cancelled 
by extra shifts of the Higgs boson fields $\phi_i^0$. 

\vspace{5mm}

{\it Acknowledgements:} 
This work was supported in part by the Grant-in-aid for 
Scientific Research from Japan Society for the Promotion of
Science, No.~12740131. 


\end{document}